\documentclass[12pt]{article}
\usepackage{amssymb}
\usepackage{amsfonts}
\usepackage{amsmath,amsthm}
\usepackage{graphics,epsfig}
\usepackage{subfigure}
\usepackage{graphicx,epsfig, color }

\oddsidemargin 10mm \numberwithin{equation}{section}
\def\be{\begin{equation}}
\def\ee{\end{equation}}
\begin{document}
\begin{center} {{\bf {The last lost charge and phase transition in Schwarzschild AdS minimally coupled to a cloud of strings}}\\
 \vskip 0.50 cm
  {{ Hossein Ghaffarnejad \footnote{E-mail: hghafarnejad@semnan.ac.ir
 } and  }{Mohammad Farsam \footnote{E-mail: mhdfarsam@semnan.ac.ir
 } }}\vskip 0.2 cm \textit{Faculty of Physics, Semnan
University, P.C. 35131-19111, Semnan, Iran }}
\end{center}

\begin{abstract}
In this paper we study the Schwarzschild AdS black hole with a cloud of string background in an extended phase space and investigate a
 new phase transition related to the topological charge. By treating the topological charge as a new charge for black hole solution we study its
  thermodynamics in this new extended phase space. We treat by two approaches to study the phase transition behavior via both $T-S$ and $P-v$
  criticality and we find the results confirm each other in a nice way. It is shown a cloud of strings affects the critical physical quantities and
   it could be observed an interesting Van der Waals-like phase transition in the extended thermodynamics. The swallow tail-like behavior is also
   observed in Free Energy-Temperature diagram. We observe in $a\to 0$ limit the small/large black hole phase transition reduces to the Hawking-Page
   phase transition as we expects. We can deduce that the impact of cloud of strings in Schwarzschild black hole can bring Van der Waals-like black hole
    phase transition.
\end{abstract}
\section{Introduction}
~~Black hole thermodynamics in Anti de-Sitter spacetime has been a
hot topic in the recent past. In a remarkable work Hawking and
Page discovered a phase transition between AdS black holes and a
global AdS space [1]. Witten explained the Hawking-Page phase
transition in terms of the AdS/CFT correspondence as a dual of the
QCD confinement/deconfinement transition [2-3]. In [4-5] Chamblin
et al found a Van der Waals like phase transition in
Reissner-Nordstr\"{o}m AdS black hole. Recently Kubiznak and Mann
discovered a surprising analogy between Reissner-Nordstr\"{o}m AdS
black holes and Van der Waals fluid-gas system in the extended
phase space of thermodynamics [6]. The extended phase space refers
to a phase space in which the first law of black holes is
corrected by a $VdP$ term and the cosmological constant is
regarded as thermodynamical pressure of the black hole and its
conjugate variable is a volume covered by the event horizon of the
black hole [7-8]. Thermodynamical behaviors of a wide range of AdS
black holes are studied in details in several works [9-27]. In a
different and interesting work, Tian [28-29] introduces the
spatial curvature of Reissner-Nordstrom as a topological charge
that naturally arises in holography. The author called it "the
last lost charge" because it is shown that topological charge
appears in an extended first law with all other known charge
(mass, electric charge, angular momentum) as a new variable and
satisfies the Gibbs-Duhem like relation. The phase transition
related to the topological charge
in Reissner-Nordstr\"{o}m AdS black hole is studied in [30-31].\\
In the other side, string theory that interprets particles as
vibration modes of one dimensional string objects can addresses
some problems of quantum gravity. A cloud of strings is also a
configuration for one dimensional strings which could be
introduced as an attractive level of activities to study the
gravitational effect
 of matter. It would be helpful considering this configuration for some reasons such as the ability of extension from four dimension to any arbitrary
  higher dimension space-time. After the prior work of Letelier [35] many authors have studied various gravitational models with different sources
   surrounded by a fluid of strings [36-40] which is analogous to a pressureless perfect fluid.
 The thermodynamics properties of the Schwarzschild AdS black hole
surrounded by a cloud of strings background in a non-extended
phase space was reported in [32]. In this paper we would like to
study the new extended phase space of Schwarzschild AdS black hole
with an energy-momentum tensor coming from a cloud of strings
related to the topological charge by two formal approaches which
leads to the same result. At the first approach we seek $T-S$
criticality behavior of phase transition around the critical point
and try to confirm it by plotting diagrams of free energy against
temperature. At the second approach we try to study these
behaviors in a $P-v$ criticality diagram.  One can see the impact
of cloud of strings can bring Van der Waals like  black hole phase
transition in an extended phase space. As we know Schwarzschild
black hole never undergoes a phase transition, but as an
interesting result in this work we see that the string cloud
background completely changes the black hole thermodynamics and
gives it a critical behavior. We also study the coexistence line
in $P-T$ diagram at which two phases are in equilibrium and ends
at the critical point. Finally the behavior of system is studied
near the critical point by calculating the critical exponent, we
conclude that the effect of string clouds and topological charge
can not change these exponents. \\ Layout of the paper is as
follows. In section 2 we define metric solution of an AdS
Schwarzschild black hole in presence of a cloud of string and
obtain the first Law of this black hole thermodynamics defined on
the horizon. In section 3 we seek its phase transition and $T-S$
and $P-V$ critically and then determine critical exponents.
Section 4 denotes to conclusion and outlook of the paper.

\section{Topologically Schwarzschild AdS minimally coupled to a cloud of strings }
~~
The action of Einstein gravity coupled to the cloud of strings can be written as
\begin{equation}
 S = \frac{1}{2}\int \sqrt{-\mathbf{g}} \big(R - 2 \Lambda ) d^{n+1}x
         + \int_{\Sigma} m \sqrt{-\mathbf{\gamma}} d\lambda^{0} d\lambda^{1},
\end{equation}
 where the last term is a Nambu-Goto action. Note that  $\mathbf{g}$ is determinant of the background metric, $m$ is a non-negative
  constant related to the string
 tension, $(\lambda^{0}, \lambda^{1})$ is a parametrization of the world sheet $\Sigma$ and $  \gamma$ is the determinant of the induced metric
\[     \gamma_{a b} = g_{\mu \nu} \frac{\partial x^{\mu}}{\partial \lambda^{a}} \frac{\partial x^{\nu}}{\partial \lambda^{b}}.     \]
 The energy momentum tensor for a cloud of strings is given by
\begin{equation}
T^{\mu \nu}=(-\gamma)^{-\frac{1}{2}}\rho \Sigma^{\mu \sigma}\Sigma_{\sigma}^{\nu},
\end{equation}
 where the number density of a string cloud is described by $\rho$ and $\Sigma^{\mu \nu}$ is the spacetime bivector
\begin{equation}
\Sigma^{\mu \nu}=\epsilon^{ab} \frac{\partial x^{\mu}}{\partial \lambda^{a}} \frac{\partial x^{\nu}}{\partial \lambda^{b}}.
\end{equation}

The black hole solution for the Einstein gravity coupled to a
cloud of strings has been derived [33] as follows.
\begin{equation}
ds^2=f(r)dt^2-f(r)^{-1}dr^2-r^{2}g_{ij}dx^{i} dx^{j}
\end{equation}
with
\begin{equation}
f(r)= k - \frac{2 m}{r^{n-2}} + \frac{r^2}{l^2} - \frac{2 a}{(n-1) r^{n-3}},
\end{equation}
where the metric function $g_{ij}$ denotes to metric function of
the $(n-1)$ dimensional hyper-surface with constant scalar
curvature $(n-1)(n-2)k$. The constant $k$ specifies the geometric
property of the hyper-surface, which takes the values $0$, $1$ and
$-1$ for flat, spherical and hyperbolic respectively. The ADM mass
$M$, the Unruh-Verlinde temperature $T$ and the Wald-Padmanabhan
entropy $S$ can be written respectively as follows.
\begin{equation}
M = k\frac{r_+^{n-2}}{4}+\frac{(n-1) r_+^n - 2 a l^2 r_+}{4 (n-1) l^2},
\end{equation}

\begin{equation}
T = \frac{ n(n-1)r_+^{n+2} + k(n-1)(n-2)l^2r_+^n - 2 a l^2 r_+^3}{4 \pi (n-1) l^2 r_+^{n+1}},
\end{equation}

\begin{equation}
S=\int^{r_+}_0\frac{1}{T}\left(\frac{\partial M}{\partial r_+}\right)dr_+=\frac{\Omega_{n-1}r_+^{n-1}}{4 G_{n+1}},
\end{equation}
where $r_{+}$ is the horizon radius and $\Omega_{n-1}$ is volume
of the $n-1$ dimensional unit sphere, as plane or hyperbola. Note
that we assume a constant for the volume of unit sphere so that in
 our study $\Omega_{n-1}=\Omega_{n-1}^{k=1}$ [27-28].

By using an equipotential surface $f(r)=constant$ and by varying
with respect to variables $k$, $r$, $m$ and $a$,  the equation
(2.8) reads

\begin{eqnarray}
  df(r,k,m,a)=4\pi T dr+1 dk-\frac{2}{r^{n-2}}dM-\frac{2}{(n-1)r^{n-3}}da=0,
\end{eqnarray}
for a zero equipotential surface. The last equation can be
rewritten as
\begin{equation}
  dM=TdS+\frac{(n-2)\Omega_{n-2}}{16\pi}r^{n-3}dk+\mathcal {A} da,
\end{equation}
for which the generalized first law become
\begin{equation}
  dM=TdS+\omega d\epsilon+\mathcal {A} da,
\end{equation}
where $ \epsilon=\Omega_{n-1}k^{\frac{n-1}{2}}$ is topological charge, $\omega=\frac{1}{8\pi}k^{\frac{3-n}{2}}r^{n-2}$ is
its conjugated potential and $\mathcal {A}=-\frac{2}{(n-1)r^{n-3}}$ is string cloud conjugated potential.

\begin{figure}[h] \centering
    \includegraphics[width=8cm,height=7cm]{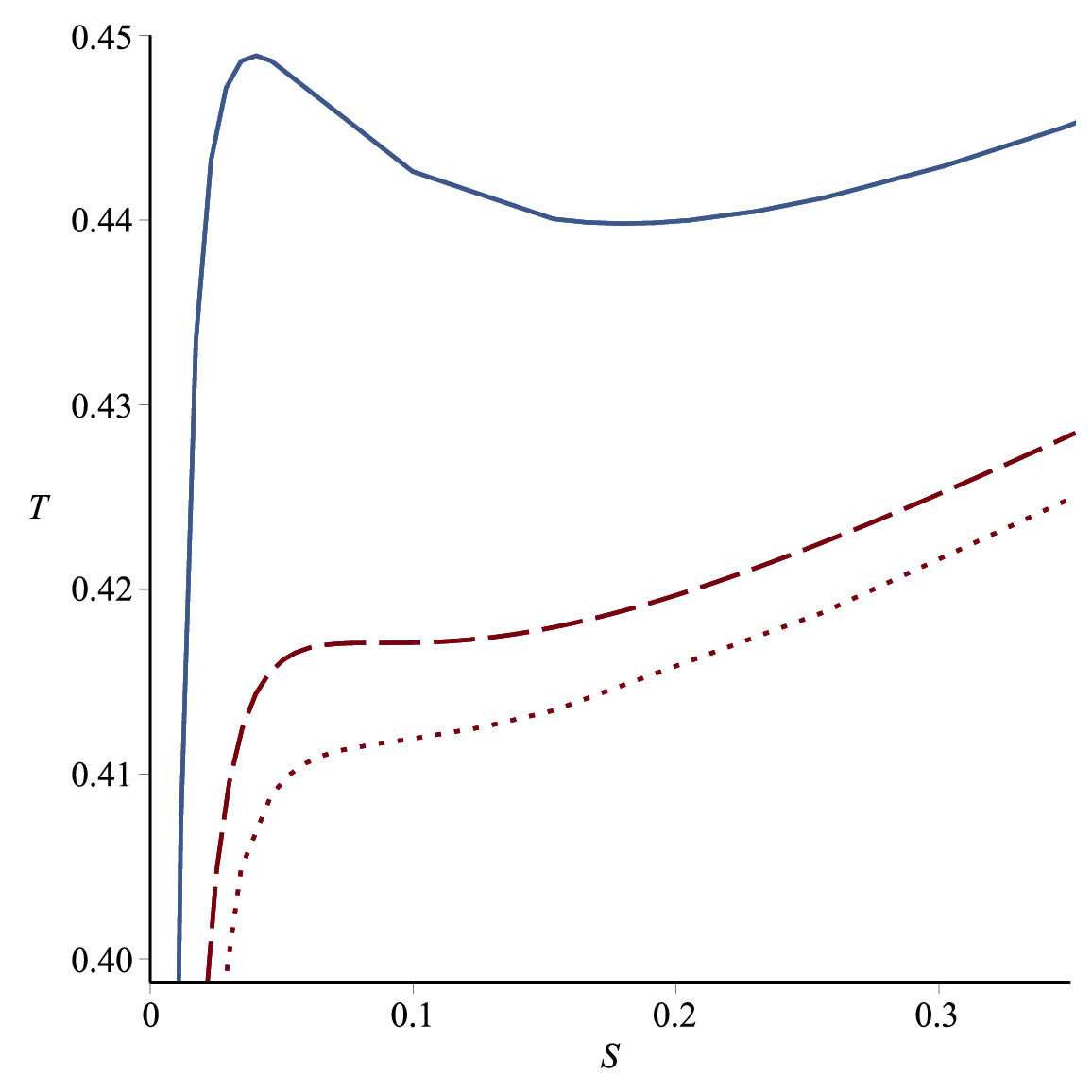}% Here is how to import EPS art
\caption{\label{fig:3}Diagram of temperature against entropy. Dot line for $\epsilon =5.03019$, dash lines for $\epsilon_{c}= 5.13019$
 and solid line for $\epsilon =5.63019$ }
\end{figure}

\begin{figure}[tbp] \centering
    \includegraphics[width=8cm,height=7cm]{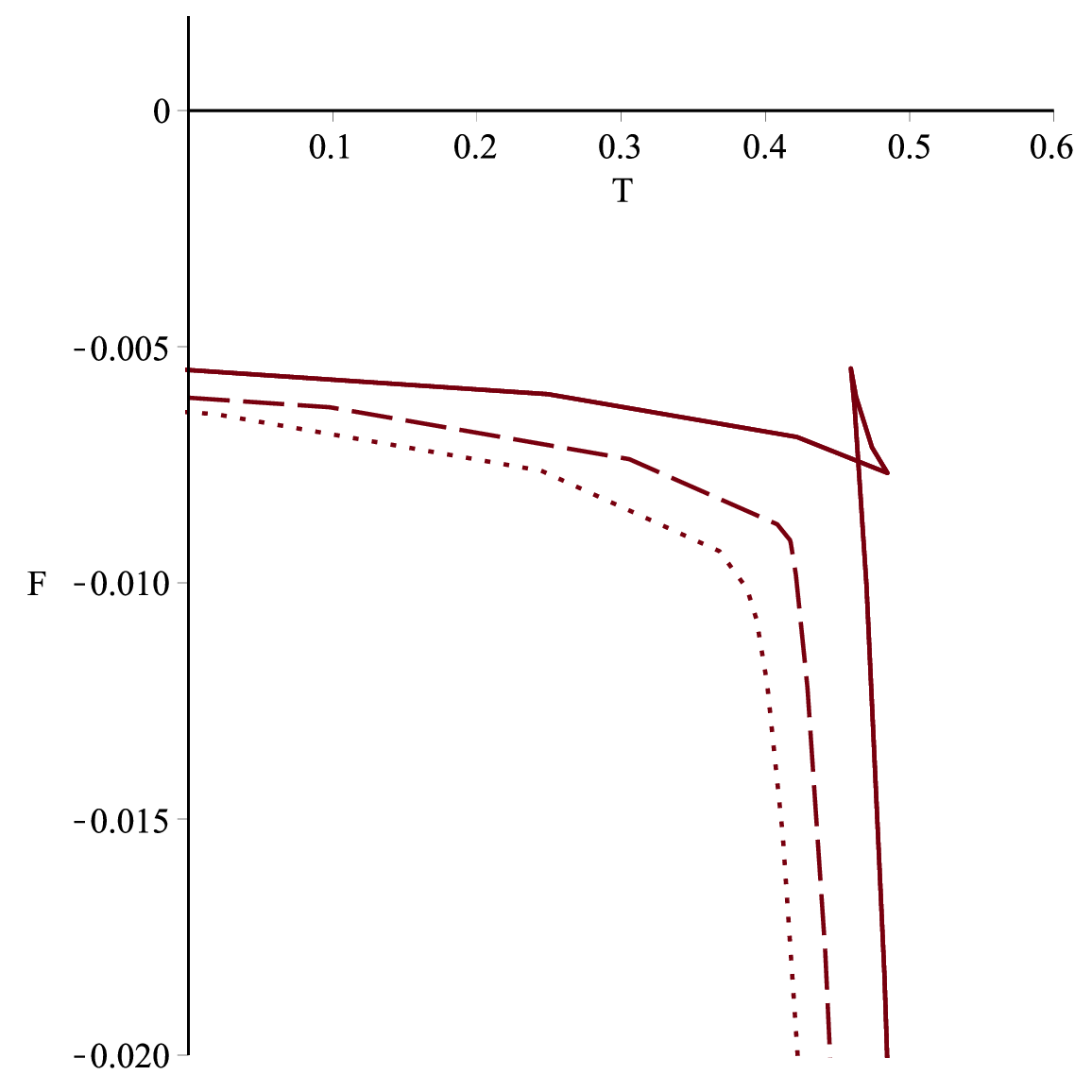}% Here is how to import EPS art
\caption{\label{fig:3}Diagram of free energy against temperature for $a=0.5$. Dot line stands for $\epsilon =4.63019$, dash lines
 for:$\epsilon_{c}= 5.13019$, and solid line for $\epsilon =6.13019$ }
\end{figure}

\begin{figure}[tbp] \centering
    \includegraphics[width=8cm,height=7cm]{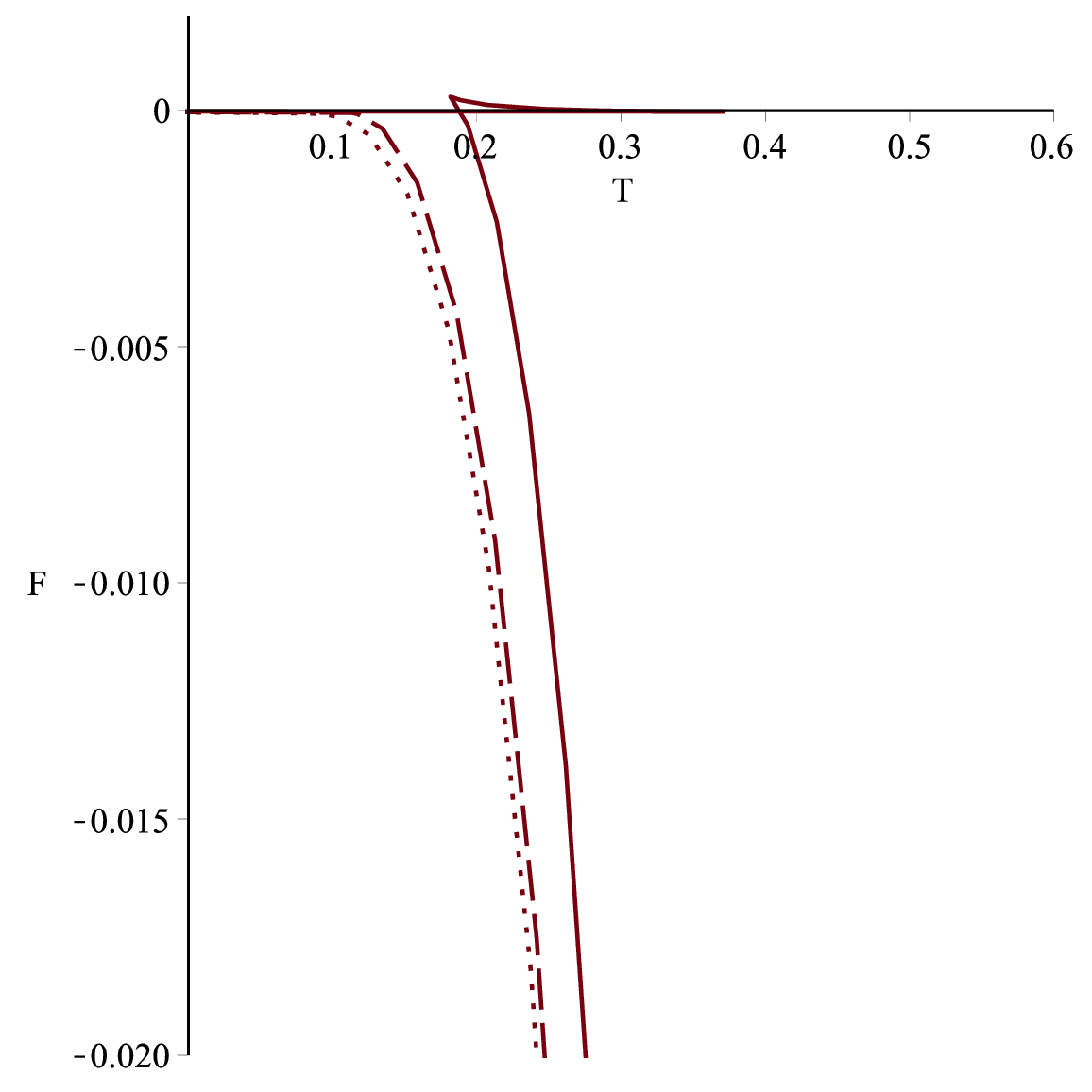}% Here is how to import EPS art
\caption{\label{fig:3}Diagram of the free energy is plotted versus
the temperature for $a=0.01$. Dot line for $\epsilon =0.07260$,
dash lines for $\epsilon_{c}= 0.10260$, and solid line: $\epsilon
=0.29260$ }
\end{figure}

\section{Critical phenomena of Schwarzschild AdS background with a cloud of strings background }
In this section, we will study the phase transition of 5
dimensional Schwarzschild AdS black hole in the presence of cloud
of strings by two approaches, $T-S$ criticality and $P-v$
criticality. Note that this solution for 4 dimensions reduces to
Schwarzschild AdS black hole which does not have Van der Waals
like phase transition (see for instance section B in ref. [8]
where by vanishing the Reissner Nordstr\"{o}m black hole electric
charge the critical point disappear).

\subsection{$T-S$ criticality}

To study phase transition and other thermodynamic behaviors of the system it must be noted that there is a
 critical hyper-surface in the parameter space and we can study it at a critical point by fixing some other parameters.
 As we can see the temperature of the black hole is a function of entropy and topological
 charge as follows.
 \begin{equation}
T=\frac{1}{4\pi}\Bigg(4\Bigg(
\frac{3S}{\pi}\Bigg)^{\frac{1}{3}}+2\Bigg(\frac{3S}{\pi}\Bigg)^{-\frac{1}{3}}\Bigg(\frac{3\epsilon}{4\pi}\Bigg)^{\frac{2}{3
}}-\frac{2}{3}a\Bigg(\frac{3S}{\pi}\Bigg)^{-\frac{2}{3}} \Bigg),
\end{equation}
for which the critical point can be calculated by
\begin{eqnarray}
\left.\frac{\partial T}{\partial S}\right|_{\epsilon=\epsilon_c}&=&0,\label{15}\\
\left.\frac{\partial ^2T}{\partial
S^2}\right|_{\epsilon=\epsilon_c}&=&0\label{16}
\end{eqnarray}
such that
\begin{equation}
S_{c}= \frac{\pi a}{18}
\end{equation}
\begin{equation}
\epsilon_{c}=10.26039    a,
\end{equation}

\begin{equation}
T_{c}=0.52551  a^{\frac{1}{3}}.
\end{equation}
The above equations show affects of the string cloud on location
of the critical points. One can see in the absence of string
cloud, criticality can not happened. On the other side the
critical points increases with string cloud factor $a$. We plot
the $T-S$ diagrams of the equation of state in figure 1. In this
figure we observe that for $\epsilon$ upper than critical
topological charge ( is shown by solid line) we have three
different solutions for the black hole's horizon which corresponds
to the identifying of them with three branches. Namely stable
large black hole, stable small black hole and an unstable medium
black hole. It is clear that below of critical topological charge
the figure is identical to the liquid-gas transition in Van der
Waals fluid. Indeed we have three cases for black hole
thermodynamic in here which are small, medium and large size black
holes. In above of critical topological charge, as it has
investigated in figure 2, medium unstable case vanishes and small
black holes is transferred to large ones straightforwardly.
Another way to investigate phase transition is studying the
behavior of free energy which is a function of temperature and
topological charges. Actually free energy has an important role in
studying thermodynamic properties of a system. Any thermodynamic
system always exists in a phase which has minimum value of free
energy among other possible values of them and a phase transition
happens when some branches of minimum free energies cross each
other. The free energy for the black hole under consideration is
given by:
\begin{equation}
F=\frac{1}{12}(kr-r^3-\frac{4}{3}a).
\end{equation}
For plotting the diagram of free energy against Unruh-Verlinde
temperature we fix $a=0.5$ and $a=0.01$ in figures 2 and 3.
 In figure 2 for topological charge lower than critical topological charge  $F-T$ diagram takes a swallow tail shape which indicates a
 first order phase transition between small and large black holes. In figure 3 we plot $F-T$ diagram for $a=0.01$ and we observe the small/large
 black hole phase transition is reduced to the Hawking-Page phase transition. In fact the effects of string cloud factor is vital to have the small/large
  black hole phase transition. So it is interesting to note that the impact of cloud of strings can bring Van der Waals like black hole phase transition.

\subsection{$P-V$ criticality}

By treating the cosmological constant as the pressure of the black
hole as
\begin{equation}
P=-\frac{\Lambda}{8\pi}=\frac{n(n-1)}{16\pi l^2},
\end{equation}
one can obtain for a 5 dimensional Schwarzschild black hole
surrounded by a cloud of string the Hawking temperature and the
generalized first law respectively as follows.
\begin{equation}
T=\frac{1}{4\pi}(4\pi vP+\frac{8}{3}\frac{k}{v}-\frac{32a}{27v^2}),
\end{equation}
and
\begin{equation}
 dM=TdS+\frac{(d-2)\Omega_{d-2}}{16\pi}r^{d-3}dk+\mathcal {A} da+V dP,
\end{equation}
where $v$ and $V$ are specific and thermodynamical volume which are defined respectively by

\begin{equation}
v=\frac{4}{3}r_+,
\end{equation}

\begin{equation}
V=\frac{\partial M}{\partial P}=\frac{1}{3}\pi r_{+}^4.
\end{equation}

Using (4.1) the black hole equation of state can be written as
\begin{equation}
P=\frac{1}{27}\frac{27\pi Tv^2-18kv+8a}{v^{3}\pi}
\end{equation}
for which the critical points can be calculated through the
conditions
\begin{eqnarray}
\left.\frac{\partial P}{\partial v}\right|_{T=T_c}&=&0,\label{15}\\,
\left.\frac{\partial ^2P}{\partial v^2}\right|_{T=T_c}&=&0.\label{16}
\end{eqnarray}
which leads to

\begin{equation}
v_c=\frac{4}{3}\frac{a}{k}
\end{equation}

\begin{equation}
P_c=\frac{1}{8\pi}\frac{k^3}{a^2}
\end{equation}
and

\begin{equation}
T_c=\frac{1}{2\pi}\frac{k^2}{a}.
\end{equation}
The above equations show effect of a cloud of strings on the
critical points where for $a \to 0$, the criticality disappears as
we expects in Schwarzschild black hole. In other words a cloud of
strings affects on criticality of the system under consideration.
The compressibility factor for our black hole solution is obtained
as $\frac{P_{c}\upsilon_{c}}{T_{c}}\simeq\frac{2.66}{8}$ which is
differ  with which one is obtained as an universal number
$\frac{P_{c}\upsilon_{c}}{T_{c}}=\frac{3}{8}$ for the Van der
Waals fluid.

\begin{figure}[h]
\centering
\subfigure[{}]{\label{1}
\includegraphics[width=.45\textwidth]{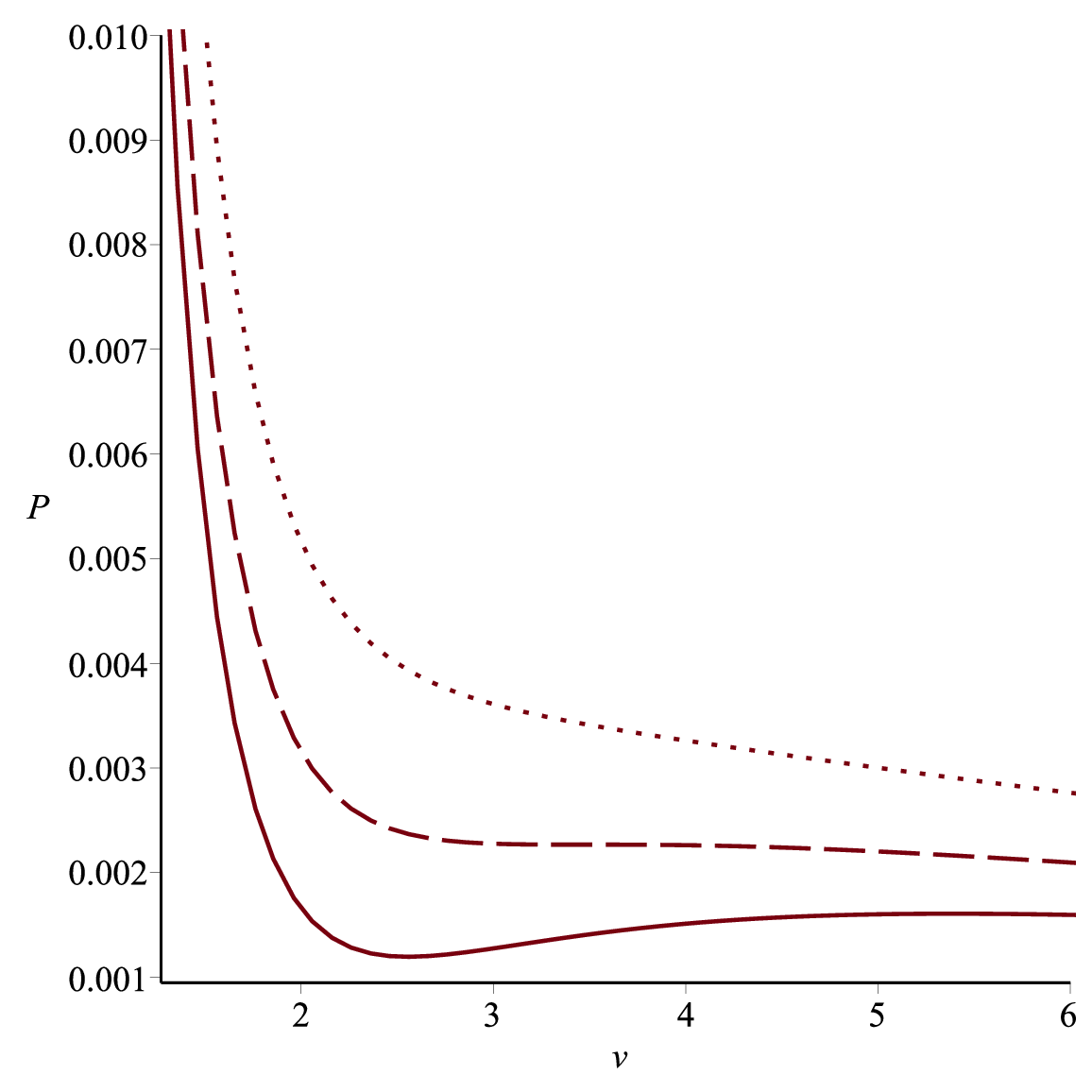}}
\hspace{1mm}
\subfigure[{}]{\label{1}
\includegraphics[width=.45\textwidth]{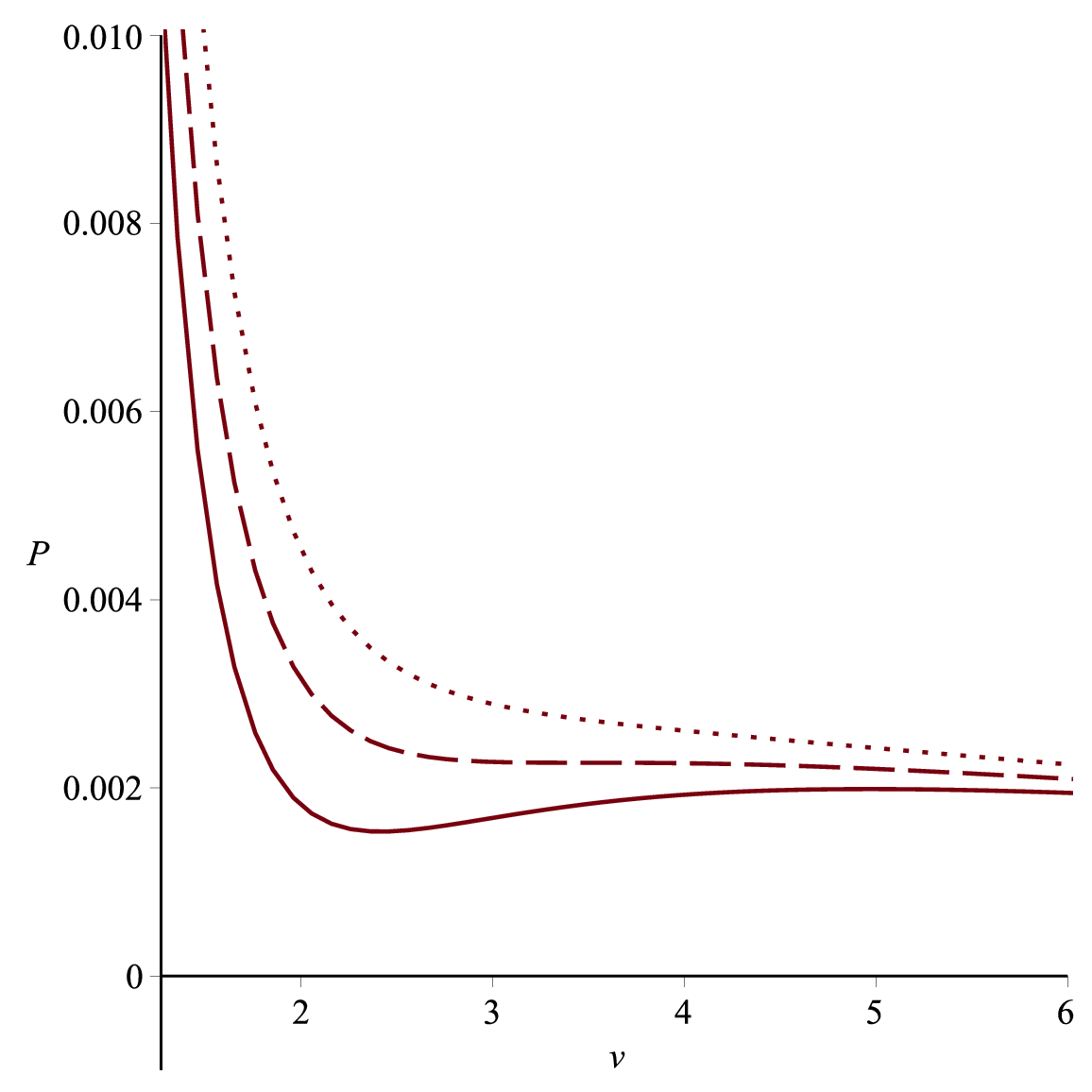}}
\caption{$P-\upsilon$ diagram in ($a$) for $a=0.1$ and $\epsilon=1$ plotted for $T_c=0.02357$, $T=0.0205<T_c$ and $T=0.0275>T_c$
represents by dash line, solid line and dot line, respectively. In diagram ($b$) it is plotted for $a=1$ and $T_c=1$, which dash line
indicates $\epsilon_c=1$, solid line $\epsilon=1.1>\epsilon_c$ and dot line $\epsilon=0.9<\epsilon_c$.}
\label{l}
\end{figure}

We plotted $P-v$ diagrams in figure 4. In these figures we plot the behavior of pressure when temperature or topological
charge is changing around their critical values. We can see from $(4.a)$ that by fixing $a$ and $\epsilon$ one can infer that the
Van der Waals-like behavior happens for temperatures lower than
the critical temperature ( in contrary with $T-S$ diagram which
criticality happens for topological charge upper than the critical
topological charge). In figure $(4.b)$ we can see by fixing $a$ and temperature the behavior is completely different.
The Van der Waals-like behavior occurs for topological charges bigger than its critical value.\\
 Applying  (3.16), (3.17), (3.18)
and defining dimensionless quantities
\begin{equation}
p=\frac{P}{P_c},~~\nu=\frac{\upsilon}{\upsilon_c},~~\tau=\frac{T}{T_c}.
\end{equation}
we can rewrite (3.13) as dimensionless form such that
\begin{equation}
p=\frac{3\tau}{\nu}-\frac{3}{\nu^2}+\frac{1}{\nu^3}.
\end{equation}
In temperature below critical temperature ($\tau<1$) there is a phase transition between small and large black hole with sizes
$\nu_s$ and $\nu_l$, respectively. In this transition the size of black hole is changed in an isobar process which could be defined by
Maxwell's equal area law, also we could see a change in the latent heat, however the Gibbs free energy keeps constant. The Gibbs free
 energy is obtained from (2.6), (2.7) and (2.8) in 5-dimension ($n=4$) and with respect to specific volume as follows:
\begin{equation}
G=M-TS=-\frac{9}{256}\pi P\upsilon^4+\frac{3}{64}k\upsilon^2-\frac{1}{12}a\upsilon,
\end{equation}
for which we can evaluate its critical value $G_c=-\frac{1}{24}\frac{a^2}{k}$ by putting $\upsilon_c$ and $P_c$ from (3.16) and (3.17),
 respectively. In figure 5 we plotted the Gibbs free energy with respect to temperature which proved the behavior we see from $P-v$ diagrams.
 Re-scaling (3.21) as $g=\frac{G}{G_c}$ would lead to
\begin{equation}
g=\frac{1}{3}p\nu^4-2\nu^2+\frac{8}{3}\nu.
\end{equation}

\begin{figure}[ht]
\centering
\subfigure[{}]{\label{1}
\includegraphics[width=.45\textwidth]{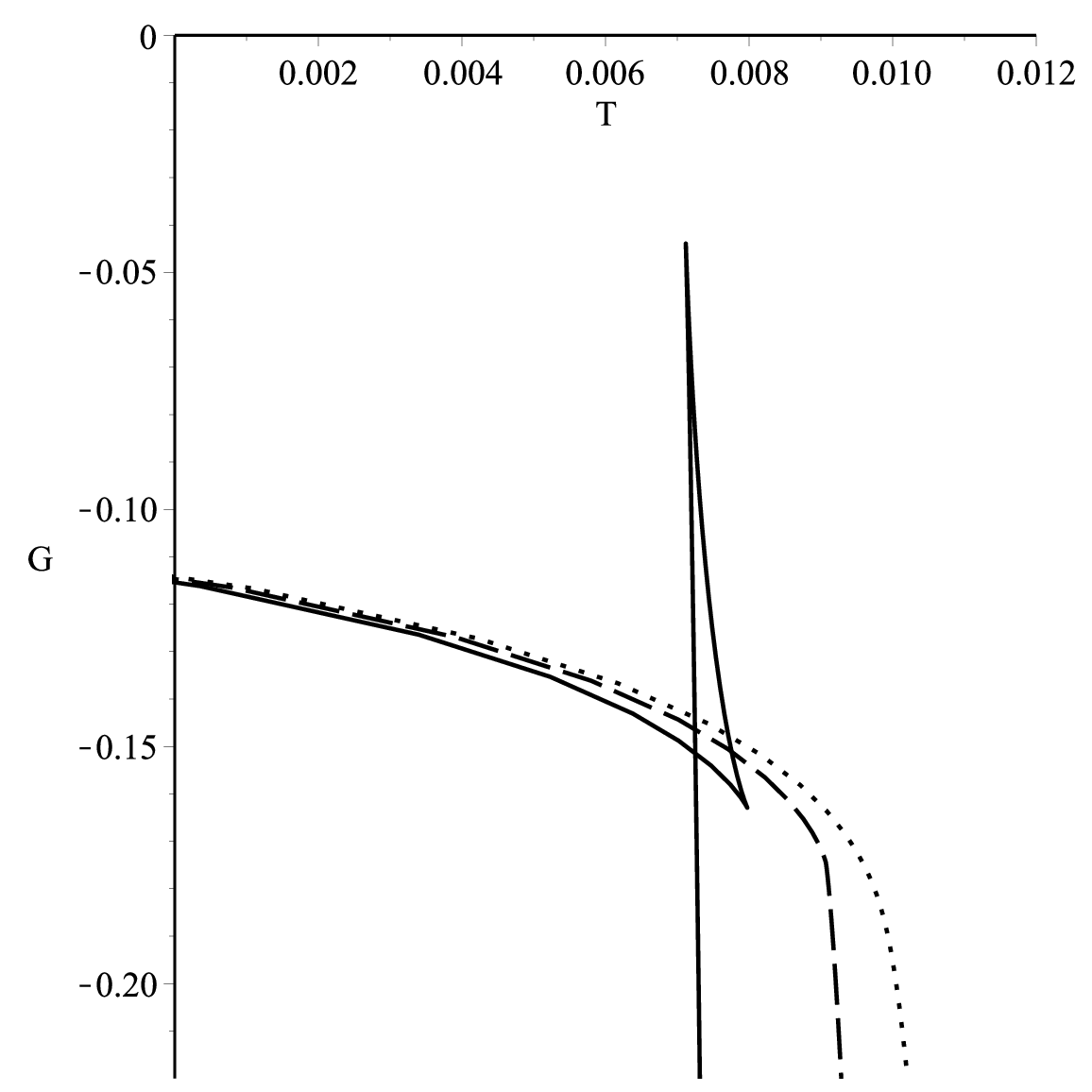}}
\hspace{1mm}
\subfigure[{}]{\label{1}
\includegraphics[width=.45\textwidth]{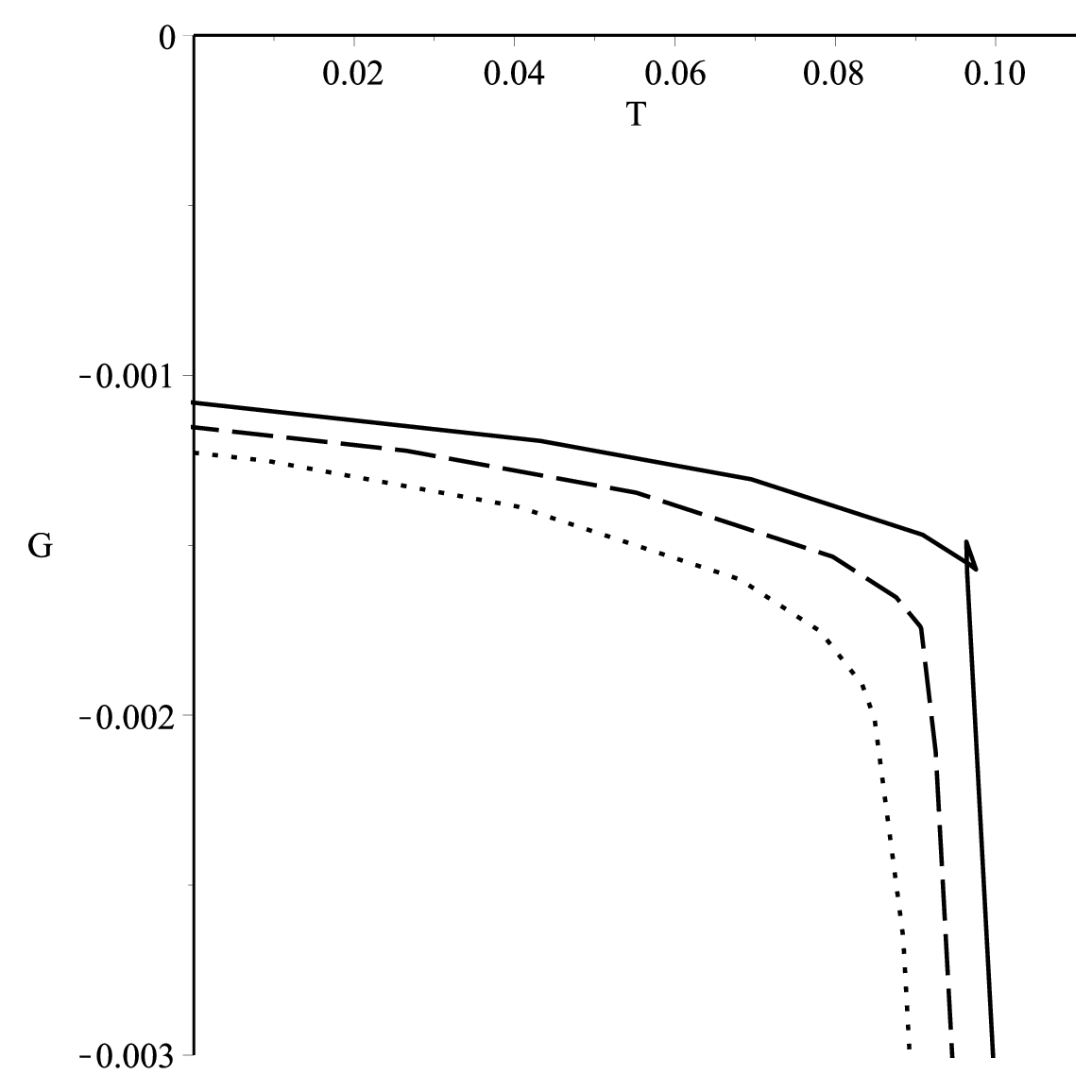}}
\caption{Gibbs free energy plotted vs temperature, diagram ($a$) for $a=1$ and $\epsilon=1$ and with $p_c=0.0005413701577$, $p=0.0003$ and
$p=0.0007$ indicated by dash line, solid line and dotted line, respectively. In $b$ it is plotted for $a=0.1$ and $p_c=0.05413701577$, with
$\epsilon_c=1$ indicated by dash line, $\epsilon=1.1$ by solid line and $\epsilon=0.9$ dotted line.  }
\label{l}
\end{figure}

It also would be helpful to study the coexistence line for which two phases are in equilibrium.
 We can find this line in $P-T$ plane for which Gibbs free energy and Hawking temperature stay constant during the transition between two
 different sizes of black hole $\nu_l$ and $\nu_s$ for large and small volume, respectively. Regarding re-scaled equations (3.20) and (3.22)
  we can find this curve as a $p-\tau$ diagram which ends at critical point by the following conditions.\begin{itemize}
             \item $g_l=g_s$:
             \begin{equation}
             \frac{1}{3}p\nu_l^4-2\nu_l^2+\frac{8}{3}\nu_l=\frac{1}{3}p\nu_s^4-2\nu_s^2+\frac{8}{3}\nu_s
             \end{equation}
             \item $\tau_l=\tau_s$:
             \begin{equation}
             \frac{p\nu_l}{3}+\frac{1}{\nu_l}-\frac{1}{3\nu_l^2}=\frac{p\nu_s}{3}+\frac{1}{\nu_s}-\frac{1}{3\nu_s^2}
             \end{equation}
             \item $2\tau=\tau_l+\tau_s$:
             \begin{equation}
             2\tau=\big(\frac{p\nu_l}{3}+\frac{1}{\nu_l}-\frac{1}{3\nu_l^2}\big)+\big(\frac{p\nu_s}{3}+\frac{1}{\nu_s}-\frac{1}{3\nu_s^2}\big)
             \end{equation}
           \end{itemize}
By plotting re-scaled pressure with respect to re-scaled temperature one can observe the coexistence line as it plotted in figure 6. As we can see
 diagram ended at re-scaled critical point $p_c=\tau_c=1$. We can see this curve
 indicates a coexistence line of small and large black hole (SBH and LBH) in $p-\tau$ plane.
\begin{figure}[tbp] \centering
    \includegraphics[width=8cm,height=7cm]{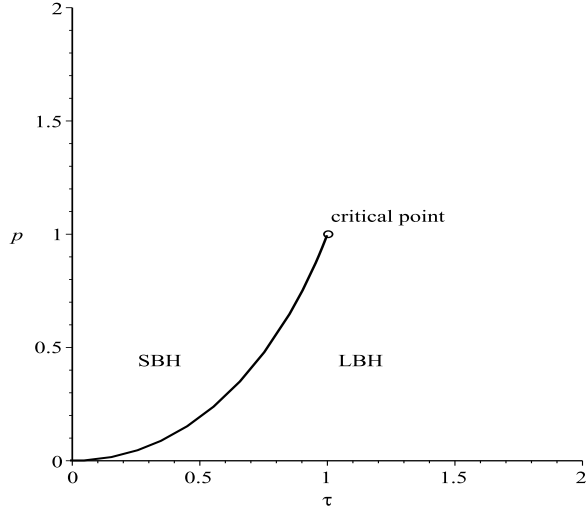}% Here is how to import EPS art
\caption{\label{fig:3} Coexistence line in $p-\tau$ plane for Schwarzschild AdS black hole in the presence of topological charge and cloud of strings. }
\end{figure}

\subsection{Behavior near the critical point}
To study the behavior of the system near the critical points, the
critical exponents ($\alpha,\beta,\gamma,\delta$) could
 be useful. These exponents are defined in a van der Waals thermodynamic system for $T<T_c$ as follows [34]
\begin{eqnarray}
 \nonumber C_\upsilon &\sim& |t|^{-\alpha} \\
\nonumber  \eta &\sim& |t|^{\beta} \\
 \nonumber \kappa_T &\sim& |t|^{-\gamma} \\
  P-P_c &\sim& |\upsilon-\upsilon_c|^{\delta},
\end{eqnarray}
in which $\upsilon$ in a gaseous system is volume per molecules of the system or specific volume and $t=\frac{T-T_c}{T_c}$. In
the above equations $C_v=T\big(\frac{\partial S}{\partial T}\big)_\upsilon$ is the specific heat at constant volume, $\eta=\upsilon_l-\upsilon_s$
 is the order parameter which is calculated in an isothermal process, $\kappa_T=-\frac{1}{\upsilon}\big(\frac{\partial\upsilon}{\partial P}\big)_T$
 is the isothermal compressibility and the last equation describes the behavior of pressure in a fixed temperature during the process corresponding to
 $T=T_c$.\\
Introducing the expansion parameters of temperature and volume around the critical point in an isobar process
\begin{equation}
\tau=1+t,~~\nu=1+\omega,
\end{equation}
one can expand the law of corresponding states as follows
\begin{equation}
p=1+3t-3\omega t-\omega^3+...
\end{equation}
The equations (3.21) could be applicable to obtain the exponents in our model:
\begin{itemize}
  \item From (2.8) one can see that the entropy is independent from Hawking temperature, so $C_\upsilon=0$ which leads to $\alpha=0$.
  \item The behavior of the order parameter depends on the changing of the specific volume in an isothermal process. This process gives us two equations:
   The first one is achieved from constant pressure during isothermal transition, namely $p_l=p_s$ which leads to
      \begin{equation}
      3t(\omega_l-\omega_s)+(\omega_l^3-\omega_s^3)=0,
      \end{equation}
      and the second one is obtained from the Maxwell's equal area law:
      \begin{equation}
      \int_{\omega_l}^{\omega_s}\omega\frac{dp}{d\omega}d\omega=0
      \end{equation}
      which by using (3.23) gives us another equation:
      \begin{equation}
      t(\omega_l^2-\omega_s^2)+\frac{1}{2}(\omega_l^4-\omega_s^4)=0.
      \end{equation}
      By solving two above equations we obtain $\omega_l=-\omega_s=\sqrt{-t}$, therefore $\eta=\upsilon_l-\upsilon_s=\upsilon_c(\omega_l-\omega_s)
      \sim\sqrt{-t}$, yielding $\beta=\frac{1}{2}$.
  \item Isothermal compressibility is derived as $\kappa_T=-\frac{1}{\upsilon}\big(\frac{\partial\upsilon}{\partial \omega}\big)\big(\frac{\partial
  \omega}{\partial P}\big)_T\sim\frac{1}{t}$, which leads to $\gamma=1$.
  \item The last exponent is described by the behavior of the critical isotherm for the pressure at $T=T_c$ which is obtained by putting $t=0$
   in (3.23) leads to $p-1\sim-\omega^3$ yielding $\delta=3$.
\end{itemize}
As we can see the effect of string clouds and topological charge
lead to a phase transition with the above critical exponents.

\section{Conclusion}
We studied the thermodynamics of topologically Schwarzschild AdS
minimally coupled in the presence of a cloud of strings in an
extended phase space. So that topological charge behaves as new
charge for black hole. The entropy of the black hole is invariant
under the existence of a cloud of strings but it is interesting to
note that the impact of cloud of strings can bring Van der Waals
like  phase transition for this  black hole. It is shown that the
iso-topological charges correspond to the topological charge less
than the critical  one which can be divided into three different
branches. Two branches which are correspond to the small and the
large black holes are maintained as stable while the medium black
hole branch is unstable. First order phase transition is observed
due to some  topology charges which are upper  than the critical
one. In figure 5 we fixed the temperature at critical value and
varying $\epsilon$ upper and lower than $1$, we see it is in
agreement with $T-S$ criticality. We also study the coexistence
line in $P-T$ diagram at which both phases are in equilibrium and
Gibbs free energy and Hawking temperature stay fixed during
transition. This line ends at critical point which is indicates in
figure 6 with respect to re-scaled variables. At last we
investigated behavior of the system close to this critical point
by studying critical exponents. We concluded that both string
clouds and topological charge have not any effect on these
exponents and they stay unchanged.
  \vskip .5cm
 \noindent
  {\bf References}\\
\begin{description}

\item[1.] S. W. Hawking, D.N. Page,~{\it {Thermodynamics of black holes in anti-de Sitter space}}, Comm. Math. Phys. {\bf 87}, 577, (1983).
\item[2.]  E. Witten,~{\it Anti-De Sitter space and holography},  Adv. Theor. Math. Phys. {\bf 2}, 253, (1998).

\item[3.] E. Witten,~{\it { Anti-de Sitter space, thermal phase transition, and confinement in gauge theories}}, Adv. Theor. Math. Phys. {\bf 2}, 505,
 (1998).
\item[4.]  A. Chamblin, R. Emparan, C. V. Johnson \& R. C. Myers,~{\it {Charged AdS black holes and catastrophic holography}}, Phys. Rev. {\bf D 60},
 064018, (1999) .
\item[5.]  A. Chamblin, R. Emparan, C. V. Johnson \& R. C. Myers,~{\it { Holography, thermodynamics, and fluctuations of charged AdS
 black holes}}, Phys. Rev. {\bf D 60}, 104026, (1999)
.
\item[6.]D. Kastor, S. Ray, \& J. Traschen,~{\it {  Enthalpy and mechanics of AdS black holes}}, Class. Quant. Grav. {\bf 26}, 195011, (2009).
.
\item[7.]  B. P. Dolan,~{\it {The cosmological constant and black hole thermodynamic potential}}, Class. Quant. Grav. {\bf 28}, 125020, (2011).

\item[8.] D. Kubiznak \& R. B. Mann,~{\it {P-V criticality of charged AdS black holes}}, JHEP 1207, 033, (2012).

\item[9.] S. Dutta, A. Jain and R. Soni,~{\it {Dyonic Black Hole and Holography}}, JHEP {\bf 2013}, 60, (2013), hep-th/1310.1748 .
\item[10.] X. X. Zeng and L. F. Li,~{\it {"Van der Waals phase transition in the framework of holography}}", hep-th/1512.08855.

\item[11.] J. Mo, G. Li, \& X. Xu, {\it {Effects of power-law Maxwell field on the Van der Waals like phase transition of higher
 dimensional dilaton black holes}}, Phys. Rev.{\bf D 93}, 084041, (2016).
\item[12.]  M. Zhang \& W Liu,l, {\it {Coexistent physics of massive black holes in the phase transitions}}, gr-qc/1610.03648.
\item[13.]  R Cai, L. Cao, \& R. Yang,, {\it {P-V criticality in the extended phase space of Gauss-Bonnet black holes in AdS space}},
 JHEP,\textbf{1309}, 005, (2013).
\item[14.]  R.  Cai, Y. Hu, Q. Pan, \& Y. Zhang, {\it { Thermodynamics of black holes in massive gravity}},
JHEP\textbf{1309}, 005, (2013).
\item[15.] J. Mo, G. Li, \& X. Xu, {\it {Combined effects of f(R) gravity and conformaly invariant Maxwell field on the extended phase
 space thermodynamics of higher-dimensional black holes}}, Eur. Phys. J. {\bf C 76}, 545, (2016).

\item[16] R. A. Hennigar and R. B. Mann, {\it {Reentrant phase transitions and van der Waals behaviour for hairy black holes ,}}, Entropy
17, 8056–8072, (2015) .

\item[17.] D. C. Zou, S. J. Zhang and B. Wang, {\it { Critical behavior of Born-Infeld AdS black holes
in the extended phase space thermodynamics}},Phys. Rev. D\textbf{
89}, 044002, (2014).

\item[18.] N. Altamirano, D. Kubiz\v{n}\'{a}k, R. Mann and Z. Sherkatghanad,
{\it {Kerr-AdS analogue of critical point and solid-liquid-gas
phase transition}}, Class, Quantum, Gravit. 31, 042001, (2013),
hep-th/1308.2672

\item[19.] J. X. Mo, X. X. Zeng, G. Q. Li, X. Jiang, W. B. Liu, {\it {A unified phase transition picture of the charged topological
black hole in Ho\v{r}ava-Lifshitz gravity, JHEP 1310,
056(2013).}}, JHEP \textbf{1310}, 056, (2013).

\item[20.] J. X. Mo and W. B. Liu, {\it {Ehrenfest scheme for $P$-$V$ criticality in the extended phase space of black holes}},Phys. Lett.
 B \textbf{727}, 336, (2013).

\item[21.]  N. Altamirano, D. Kubiz\v{n}\'{a}k and R. Mann, {\it {Reentrant phase transitions in rotating AdS black holes}},Phys. Rev. D \textbf{88},
101502, (2013).

\item[22.] R. Zhao, H. H. Zhao, M. S. Ma and L. C. Zhang, {\it {On the critical phenomena and thermodynamics of charged topological
dilaton AdS black holes}}, Eur. Phys. J. C \textbf{73}, 2645,
(2013).

\item[23.] X.-X. Zeng, X.-M. Liu and L.-F. Li, {\it {Phase structure of the Born-Infeld-anti-de Sitter black
holes probed by non-local observables}},Eur. Phys. J.C\textbf{76},
616, (2016).

\item[24.] H. Liu and X.-h. Meng, {\it {PV criticality in the extended phasespace of charged accelerating
AdS black holes}}Mod. Phys. Lett.A\textbf{31}, 1650199, (2016) .

\item[25] D. Hansen, D. Kubiznak and R. B. Mann, {\it {Universality of P-V Criticality in Horizon Thermodynamics}},
 JHEP \textbf{01}, 047, (2017), gr-qc/1603.05689.

\item[26] A. Rajagopal, D. Kubiznak and R. B. Mann, {\it {an der Waals black hole}}, Phys. Lett. B \textbf{737}, 277, (2014).

\item[27] Y. Tian, X.N. Wu, H. Zhang, {\it {Holographic Entropy Production}}, JHEP \textbf{10},  170 , (2014).

\item[28] Y. Tian, {\it {the last lost charge of a black hole}}, gr-qc/1804.00249.

\item[29] Shan-Quan Lan, Gu-Qiang Li, Jie-Xiong Mo, Xiao-Bao Xu, {\it {A New Phase Transition Related to the Black
Hole's Topological Charge}}, gr-qc/1804.06652, (2018).

\item[30] Ghaffarnejad H, Yaraie E, Farsam M. {\it {Quintessence Reissner Nordström anti de Sitter black holes and Joule
Thomson effect.}},Int. J. Theor. Phys.57, 1671, (2018).

\item[31] J. P. Morais Grac§a, Iarley P. Lobo, I. G. Salako, {\it {Cloud of strings in f(R) gravity}}, Chinese Physics C \textbf{42}, 063105
 (2018),  gr-qc/1708.08398.

\item[32] T. K. Dey, {\it {Thermodynamics of AdS Schwarzshild black hole in the presence of external string cloud}}, hep-th/1711.07008 (2018).
\item[33] S. G. Ghosh, U. Papnoi and S. D. Maharaj, {\it {Cloud of strings in third order Lovelock gravity}} Phys. Rev. D. \textbf{90}, 044068, (2014) .

\item[34.] Goldenfeld, Nigel. Lectures on phase transitions and the renormalization group. CRC Press, 2018.
\item[35.] Letelier, Patricio S. "Clouds of strings in general relativity." Physical Review D \textbf{20},1294, (1979).
\item[36.] Ghosh, Sushant G., and Sunil D. Maharaj. "Cloud of strings for radiating black holes in Lovelock gravity." Phys. Rev.
 D \textbf{89}, 084027, (2014).
\item[37.] Ghosh, Sushant G., Uma Papnoi, and Sunil D. Maharaj. "Cloud of strings in third order Lovelock gravity." Phys. Rev. D
\textbf{90}, 044068, (2014).
\item[38.] Mazharimousavi, S. Habib, and M. Halilsoy. "Cloud of strings as source in $2+1$-dimensional $f(R)= R^{n}$ gravity." The
Eur. Phys. J. C 76, 95, (2016).
\item[39.] Toledo, Jefferson de M., and V. B. Bezerra. "Black holes with cloud of strings and quintessence in Lovelock gravity."
 The Eur. Phys. J. C 78, 534, (2018).
\item[40.] Ganguly, Apratim, Sushant G. Ghosh, and Sunil D. Maharaj. "Accretion onto a black hole in a string cloud background." Phys. Rev. D
\textbf{90},  064037 (2014).

\end{description}

\end{document}